\documentclass[twocolumn,aps,showpacs,prb]{revtex4-1}
\usepackage{graphicx}
\usepackage{epstopdf}
\usepackage{amsmath}
\usepackage{multirow}

\begin{document}

\title{Tuning the magnetism of the top-layer FeAs on BaFe$_{2}$As$_{2}$(001): First-principles study}
\author{Bing-Jing Zhang$^{1,2}$}
\author{Kai Liu$^{1,2}$}\email{kliu@ruc.edu.cn}
\author{Zhong-Yi Lu$^{1,2}$}\email{zlu@ruc.edu.cn}

\affiliation{$^{1}$Department of Physics, Renmin University of China, Beijing 100872, China}
\affiliation{$^{2}$Beijing Key Laboratory of Opto-electronic Functional Materials $\&$ Micro-nano Devices, Beijing 100872, China}

\date{\today}

\begin{abstract}

The magnetic properties of BaFe$_{2}$As$_{2}$(001) surface have been studied by using first-principles electronic structure calculations. We find that for As-terminated surface the magnetic ground state of the top-layer FeAs is in the staggered dimer antiferromagnetic (AFM) order, while for Ba-terminated surface the collinear (single stripe) AFM order is the most stable. When a certain coverage of Ba or K atoms are deposited onto the As-terminated surface, the calculated energy differences among different AFM orders for the top-layer FeAs on BaFe$_{2}$As$_{2}$(001) can be much reduced, indicating enhanced spin fluctuations. To identify the novel staggered dimer AFM order for the As termination, we have simulated the scanning tunneling microscopy (STM) image for this state, which shows a different $\sqrt{2}\times\sqrt{2}$ pattern from the case of half Ba coverage. Our results suggest: i) the magnetic properties of the top-layer FeAs on BaFe$_{2}$As$_{2}$(001) can be tuned effectively by surface doping; ii) both the surface termination and the AFM order in the top-layer FeAs can affect the STM image of BaFe$_{2}$As$_{2}$(001).

\end{abstract}


\maketitle

\section{INTRODUCTION}

Investigating the magnetic properties of iron-based superconductors is very crucial for understanding the pairing mechanism of their unconventional superconductivity\cite{1Scalapino(2012)RMP}. Most parent compounds of iron-based superconductors show long range magnetic order at low temperature, while their superconductivity can be induced by suppressing the magnetic order via charge doping or pressure. Among various iron-based superconductors, the '122' compound BaFe$_{2}$As$_{2}$ is widely studied due to its high quality of single crystal\cite{2Johnston(2010)AIP,3GStewart(2011)RMP}. The magnetic ground state of BaFe$_{2}$As$_{2}$ is in the collinear (single stripe) antiferromagnetic (AFM) order \cite{4PDai(2012)NP}, and the superconductivity can be induced by hole doping\cite{5HChen(2009)EPL} with K replacing Ba or electron doping\cite{6(Codoping)NandiS(2010)PRL,7(Nidoping)LiLJ(2009)NJP} with Co/Ni replacing Fe, as well as by chemical pressure\cite{8SJiang(2009)JPCM} with P replacing As or external physical pressure \cite{9ColombierE(2009)PRB,10Ishikawa(2009)PRB,11KimberS(2009)NM}. It has been commonly accepted that suppressing magnetic order and enhancing spin fluctuations are responsible for the appearance of superconductivity in iron-based superconductors\cite{12TImai(2009)PRL}.

In addition to bulk crystals, the surfaces of iron-based superconductors offer a good platform for studying the underlying physical mechanism of their superconductivity. Unlike those bulk-probing tools such as electric transport and magnetic susceptibility measurements, scanning tunneling microscopy (STM) is surface-sensitive and has advantage in studying the surfaces of iron-based superconductors.  Nevertheless, it is very important to clarify whether or not the observations obtained from this surface-sensitive measurement can reflect the bulk properties. Different from a '11' compound like FeSe which has definite cleavage surface\cite{13MasseeF(2009)PRB}, a '122' compound suffers from controversial identifications of the cleavage surfaces and shows complex surface reconstructions depending on cleaving process and temperature history\cite{15NascimentoVB(2009)PRL,Li2012PRB,18NiestemskiFC(2009)arXiv:0906.2761,19ChuangTM(2010)Science,20BoyerMC(2008)arXiv:0806.4400,21YinY(2009)PRL,22MasseeF2(2009)PRB, 23MasseeF(2010)EPL,24PanSHBAPSD,25HuiZhang(2010)PRB,26VanHeumenE(2011)PRL}. For the '122' compound BaFe$_{2-x}$Co$_{x}$As$_{2}$, both the $\sqrt{2}\times\sqrt{2}$ pattern \cite{13MasseeF(2009)PRB} and the 1$\times$2 stripe pattern\cite{21YinY(2009)PRL} have been observed, meanwhile the complete As surface, the half Ba layer, and the full Ba layer have all been proposed as possible surface terminations \cite{30MGao(2010)PRB,31GProfeta(2010)PRB}. While most studies have considered the structural and electronic properties of surface terminations, little attention has been paid to the magnetic properties at the BaFe$_{2}$As$_{2}$ surface \cite{Li2012PRB}. Given the facts that K-doping in the Ba layer can effectively suppress the magnetic order in FeAs layer and induce the superconductivity in BaFe$_{2}$As$_{2}$ \cite{5HChen(2009)EPL,33HHWen(2011)RCMP} and that the magnetic properties of atomic FeSe films on SrTiO$_{3}$ can be influenced by electron doping in our previous study \cite{34Kai(2015)PRB}, it is natural to ask whether the magnetic properties of the BaFe$_{2}$As$_{2}$ surface are distinct from its bulk counterpart, since the surface suffers a very different chemical environment.

In this article, we focus on how surface environment affects the magnetism of the top-layer FeAs on BaFe$_{2}$As$_{2}$(001). In turn, we also find that the magnetic state of the top-layer FeAs, in addition to previously noticed termination structures, can have influence on the STM image of the cleavage surface. The rest of the paper is organized as follows. In section \uppercase\expandafter{\romannumeral2} , the computational details are described. Section \uppercase\expandafter{\romannumeral3} presents the calculation results and corresponding analysis. Section \uppercase\expandafter{\romannumeral4} gives the discussions of our results with related experimental and theoretical works as well as a short summary.

\section{COMPUTATIONAL DETAILS}

To investigate the electronic and magnetic properties of BaFe$_{2}$As$_{2}$(001) surface, we carried out first-principles electronic structure calculations with the projector augmented wave (PAW) method \cite{35PAW} as implemented in the Vienna \textit{ab initio} simulation package \cite{36VASP1,37VASP2}. The generalized gradient approximation (GGA) of Perdew-Burke-Ernzerhof (PBE) type was chosen for the exchange-correlation functional \cite{38PBE}. The plane-wave basis set was employed with a kinetic energy cutoff of 350 eV. The Fermi level was broadened by a Gaussian smearing method with a width of 0.05 eV. We first obtained the fully relaxed lattice parameters of bulk BaFe$_{2}$As$_{2}$ in the collinear (single stripe) AFM order, and then used its in-plane lattice constants for surface. To study different cleavage surfaces of BaFe$_{2}$As$_{2}$(001), we built a slab model with three FeAs layers and three or four Ba layers for As or Ba termination, respectively. In all cases, a vacuum layer thicker than 15 \AA\ was adopted. For the Brillouin zone sampling of these two-dimensional supercells, we adopted $4\times8\times1$ ${\bf k}$-point mesh for $2\sqrt{2}\times\sqrt{2}$ supercell and $4\times4\times1$ ${\bf k}$-point mesh for $2\sqrt{2}\times2\sqrt{2}$ supercell. Only atoms in the top three/four layers of slabs were allowed to relax until the corresponding forces were smaller than 0.01 eV/\AA, while the bottom layers were fixed at their corresponding bulk positions. The effect of electric field caused by the asymmetric slab relaxation was canceled by a dipole correction \cite{39JNeugebauer(1992)PRB}. The STM images were simulated at a height of 3 \AA\ away from the surfaces.

\begin{figure}[!t]
\includegraphics[angle=0,scale=0.4,clip=true]{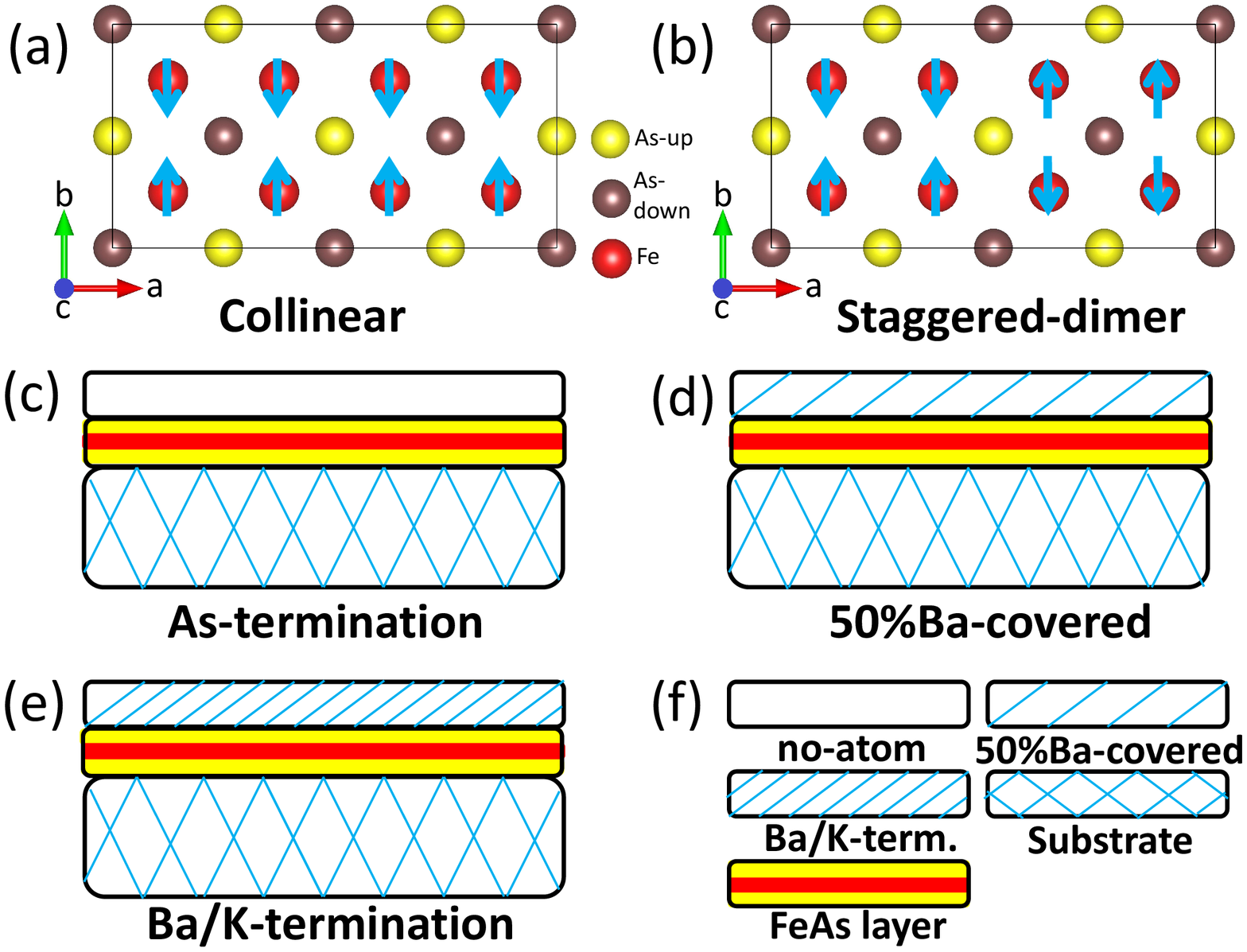}
\caption{(Color online) Top views of the spin patterns for (a) the collinear (single stripe) and (b) the staggered dimer AFM orders of BaFe$_{2}$As$_{2}$. The red balls denote Fe atoms while the yellow/brown balls represent As atoms located above/below the Fe plane. The blue arrows indicate the spin polarization directions. Schematic illustration of (c) the As-terminated, (d) the 50\%Ba-covered, and (e) the Ba- or K-terminated BaFe$_{2}$As$_{2}$(001) surfaces, respectively. Here the 50\%Ba-covered surface means that half of surface Ba atoms are left after cleavage. (f) Illustration of different blocks in the simulated slabs of BaFe$_{2}$As$_{2}$(001) surface.}
\label{fig1}
\end{figure}

\section{RESULTS AND ANALYSIS}

We have considered four types of BaFe$_{2}$As$_{2}$(001) surfaces, three of which are of the As termination [Fig. 1(c)], the 50\%Ba-covered (half-Ba-coverage) termination [Fig. 1(d)], and the Ba termination [Fig. 1(e)], respectively, as suggested by experiments \cite{13MasseeF(2009)PRB,15NascimentoVB(2009)PRL,22MasseeF2(2009)PRB,25HuiZhang(2010)PRB}. In addition, we have also simulated a proposed full-K-coverage termination [Fig. 1(e)], since it gives nominally the same amount of electron doping to the top-layer FeAs as in the 50\%Ba-covered case. For spin patterns, the Fe atoms in all inner FeAs layers of BaFe$_{2}$As$_{2}$(001) slab adopt the well-known collinear (single stripe) AFM order [Fig. 1(a)], the same as the case of the magnetic ground state in bulk BaFe$_{2}$As$_{2}$, while the Fe atoms in the top-layer FeAs may adopt different magnetic orders, i.e., possibly the collinear (single stripe) AFM order [Fig. 1(a)], the checkerboard AFM N\'eel order (not shown), or the staggered dimer AFM order [Fig. 1(b)].

On the 50\%Ba-covered BaFe$_{2}$As$_{2}$(001) surface, Ba atoms may arrange randomly or regularly, forming miscellaneous surface structures \cite{13MasseeF(2009)PRB}. Among them, the ${1}\times{2}$ stripe [Fig. 2(a)], the $\sqrt{2}\times\sqrt{2}$ square [Fig. 2(b)], and the zigzag stripe [Fig. 2(c)] structures are three typical ordering patterns for surface Ba atoms.

\begin{figure}[!t]
\includegraphics[angle=0,scale=0.365,clip=true]{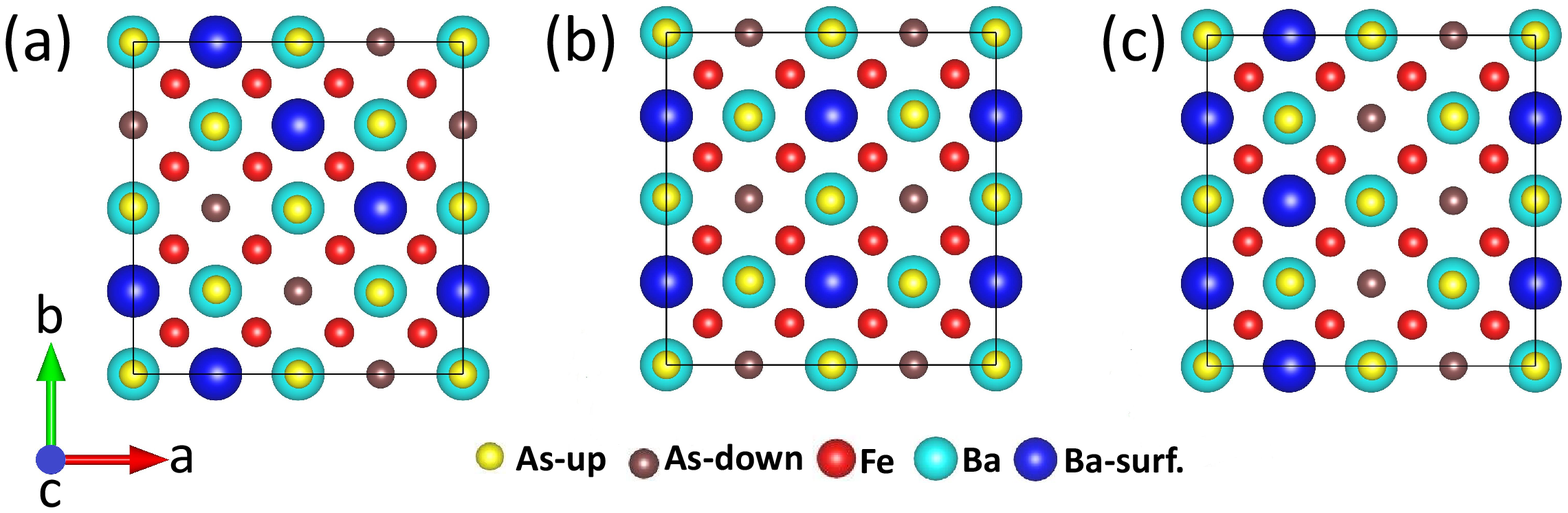}
\caption{(Color online) Three typical patterns of 50\%Ba-covered BaFe$_{2}$As$_{2}$(001) surface: (a) the $1\times2$ stripe, (b) the $\sqrt{2}\times\sqrt{2}$ square, and (c) the zigzag stripe patterns of surface Ba atoms, respectively.}
\label{fig2}
\end{figure}

\begin{table}[b!]
\caption{Relative energies (in unit of eV/supercell) for different AFM orders of the top-layer FeAs on three 50\%Ba-covered BaFe$_{2}$As$_{2}$(001) surfaces. The surface Ba atoms adopt the $1\times2$ stripe, the $\sqrt{2}\times\sqrt{2}$ square, and the zigzag stripe patterns, respectively (Fig. 2). All inner FeAs layers are in the collinear AFM order. The energy of 50\%Ba-covered BaFe$_{2}$As$_{2}$(001) with the top-layer FeAs in the AFM N\'eel order and the surface Ba atoms adopting the $1\times2$ stripe pattern is set to zero.}
\begin{center}
\begin{tabular*}{8cm}{@{\extracolsep{\fill}}cccc}
\hline
\hline
 & 1$\times$2 & $\sqrt{2}\times\sqrt{2}$ & zigzag \\
\hline
collinear & -0.790  & -1.426 & -0.927 \\
dimer & -0.456 & -1.216 & -0.731  \\
N\'eel & 0 & -0.921 & -0.284 \\
\hline
\hline
\end{tabular*}
\end{center}
\end{table}

To determine which kind of 50\%Ba-covered pattern is the most stable one, we have calculated the total energies of the $1\times2$ stripe, the $\sqrt{2}\times\sqrt{2}$ square, and the zigzag stripe patterns with collinear AFM, staggered dimer AFM, and AFM N\'eel orders in the top-layer FeAs, respectively. Table I lists the relative energies of these states. We see that for each surface-Ba pattern (in columns), the collinear AFM order always has lower energy than both the staggered dimer AFM and the AFM N\'eel orders. Meanwhile, no matter which AFM order the top-layer FeAs adopts (in rows), the system with surface Ba in the $\sqrt{2}\times\sqrt{2}$ square pattern invariably owns the lowest energy among three surface structures. In other words, the $\sqrt{2}\times\sqrt{2}$ square pattern of surface Ba atoms with the top-layer FeAs in the collinear AFM order is the magnetic ground state of the 50\%Ba-covered BaFe$_{2}$As$_{2}$(001) surface. Hereafter, we use this surface structure for the 50\%Ba-covered case to compare with other terminations.

Figure 3 shows the relative energies of the staggered dimer AFM order as well as the AFM N\'eel order with respect to the collinear AFM order for the top-layer FeAs on BaFe$_{2}$As$_{2}$(001) at different terminations, respectively. It can be seen that for the As termination, the energy of the staggered dimer AFM order is the lowest, indicating this AFM order being of the magnetic ground state. With more Ba or K atoms deposited onto the As-terminated surface, there are more electrons doped into the top-layer FeAs. It turns out that the energy difference between the staggered dimer and the collinear AFM orders becomes almost zero at the full K coverage and then enlarges gradually at the half and full Ba coverages. Meanwhile, the energy of the AFM N\'eel order rises up quickly and exceeds those of other two AFM orders. Thus when a certain coverage of Ba or K atoms are deposited onto the As-terminated surface, the energy differences between different AFM orders can be much reduced, while at a higher Ba coverage with more electron doping, the magnetic ground state of the top-layer FeAs recovers to its bulk counterpart, i.e., to the collinear AFM order. These results indicate that surface doping can effectively tune the magnetic properties of the top-layer FeAs on BaFe$_{2}$As$_{2}$(001).

\begin{figure}[!t]
\includegraphics[angle=0,scale=0.35,clip=true]{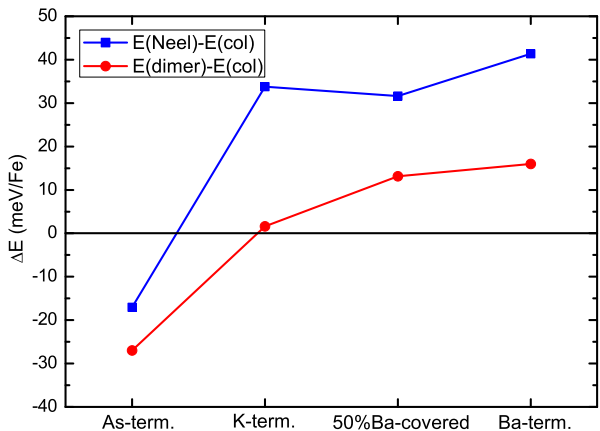}
\caption{(Color online) The variations in the energy differences between the staggered dimer and the collinear AFM orders, as well as those between the AFM N\'eel and the collinear AFM orders, of the top-layer FeAs on BaFe$_{2}$As$_{2}$(001) at different terminations. For the 50\%Ba-covered BaFe$_{2}$As$_{2}$(001), the surface Ba atoms form a $\sqrt{2}\times\sqrt{2}$ square pattern as in Fig. 2(b).}
\label{fig3}
\end{figure}

Previously, a computational study found that in bulk FeSe the calculated energy of the staggered dimer AFM order is also lower than that of the collinear AFM order \cite{40HYCao(2015)PRB}. Further calculations showed that the magnetic frustration \cite{41JKGlasbrenner(2015)NP}, more specifically, the energetically almost degenerate staggered dimer and staggered trimer AFM orders as well as their random combinations \cite{42Kai(2016)PRB}, suppress the static magnetic order of bulk FeSe. Here, for the As-terminated BaFe$_{2}$As$_{2}$(001) surface, our calculations show that the energy of the staggered dimer AFM order of the top-layer FeAs is about 8 meV/Fe lower than that of the staggered trimer AFM order, indicating the former may be observable in experiment.

\begin{figure}[!t]
\includegraphics[angle=0,scale=0.325,clip=true]{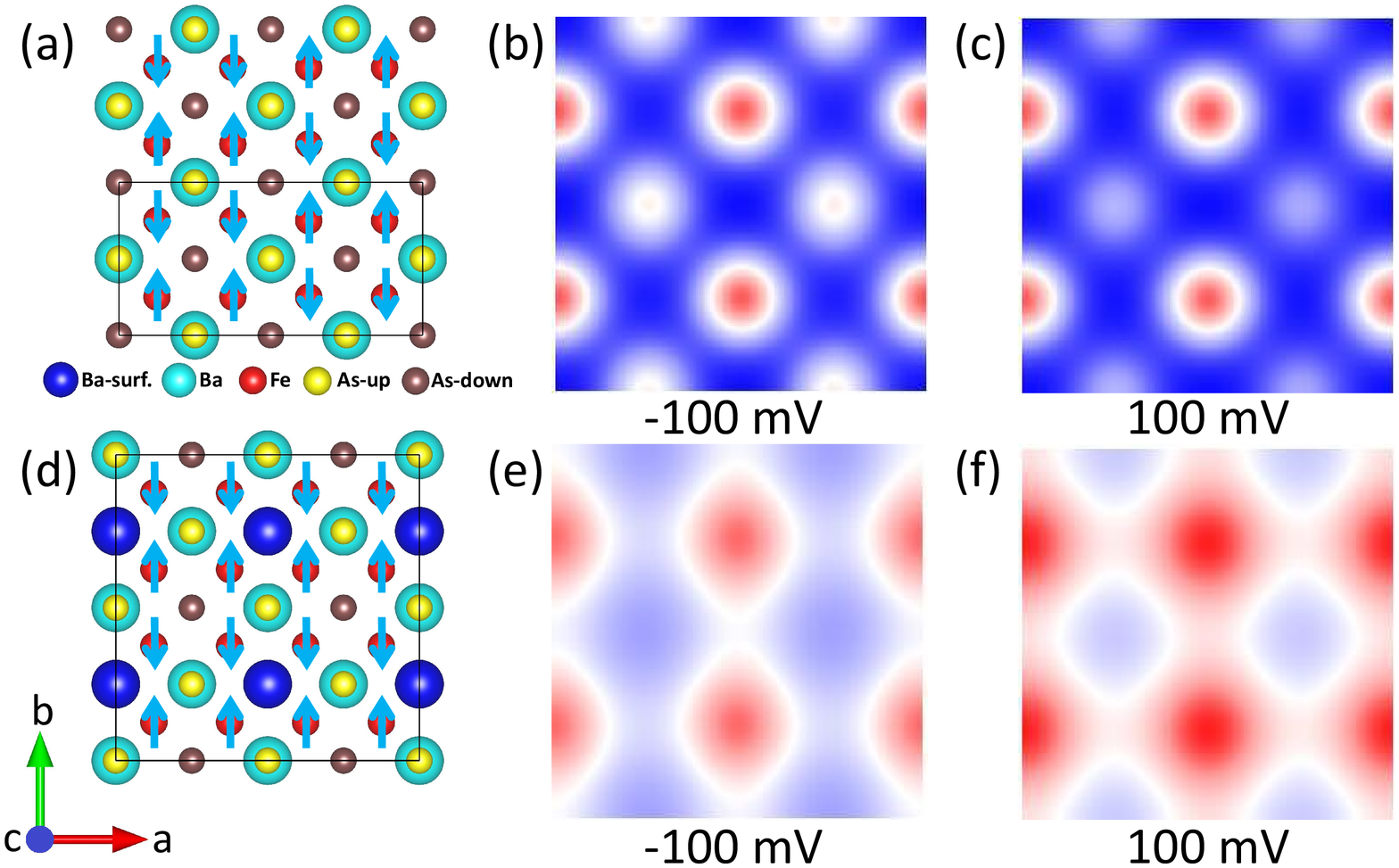}
\caption{(Color online) The structures and the simulated STM images of different BaFe$_{2}$As$_{2}$(001) surfaces at -100 and 100 mV biases. (a)-(c) The As-terminated surface with the top-layer FeAs in the staggered dimer AFM order. (d)-(f) The 50\%Ba-covered surface with the surface Ba atoms adopting the $\sqrt{2}\times\sqrt{2}$ square pattern [Fig. 2(b)] and the top-layer FeAs in the collinear AFM order.}
\label{fig4}
\end{figure}

Figures 4(b) and 4(c) display the simulated STM images of the As-terminated BaFe$_{2}$As$_{2}$(001) surface under different biases with the top-layer FeAs in the staggered dimer AFM order. No matter whether a negative or positive bias is imposed, namely corresponding respectively to the occupied or unoccupied states, the STM morphologies always show a clear $\sqrt{2}\times\sqrt{2}$ periodicity formed by red or white spots. The red spots are located well above one set of surface As atoms, while the white spots sit above the other set, ascribing to different environments around these two sets of surface As atoms created by the underlying staggered dimer Fe spins [Fig. 4(a)]. On the other hand, previous experiment showed that the 50\%Ba-covered BaFe$_{2}$As$_{2}$(001) surface also displays a $\sqrt{2}\times\sqrt{2}$ pattern \cite{25HuiZhang(2010)PRB}. We have thus studied the STM image of the magnetic ground state of the 50\%Ba-covered BaFe$_{2}$As$_{2}$(001) surface (Table I), for which an apparent $\sqrt{2}\times\sqrt{2}$ pattern formed by the bright diffuse spots can be discerned [Figs. 4(e) and 4(f)]. These spots are located right above the surface Ba atoms. Nevertheless, different from the As-terminated case [Figs. 4(b) and 4(c)], the 50\%Ba-covered termination only shows one set of $\sqrt{2}\times\sqrt{2}$ grid [Figs. 4(e) and 4(f)]. We have also simulated the STM images for other terminations (see Fig. 5 in the Appendix), yet none of them shows a $\sqrt{2}\times\sqrt{2}$ pattern. Thus the simulated STM images here provide typical characteristics to identify the staggered dimer AFM order of the top-layer FeAs on the As-terminated BaFe$_{2}$As$_{2}$(001).

\section{DISCUSSION AND SUMMARY}

The above calculations demonstrate that the magnetic order in ground state of the top-layer FeAs on BaFe$_{2}$As$_{2}$(001) varies with different terminations: for the As termination it is the staggered dimer AFM order, while for the half- or full- coverage Ba termination it recovers to its bulk counterpart, i.e., the collinear (single stripe) AFM order (Fig. 3). According to the Heisenberg spin model, different magnetic ground states result from the interplays among the nearest-neighbor exchange interaction ($J_1$), the next-nearest-neighbor exchange interaction ($J_2$), and the next-next-nearest-neighbor exchange interaction ($J_3$) of Fe spins bridged by anion atoms (an effective $J_1$-$J_2$-$J_3$ Heisenberg model) \cite{Ma08,Ma09}. When $J_1-2J_2+2J_3>0$, the staggered dimer AFM order \cite{40HYCao(2015)PRB} and the staggered $n$-mer ($n$=2, 3, ...) AFM order \cite{42Kai(2016)PRB} will be energetically lower than the collinear AFM order. For bulk FeSe, the energy difference between the staggered dimer and the staggered trimer AFM orders is negligible \cite{42Kai(2016)PRB}, which leads to the absence of static magnetic order in bulk FeSe. Here, the predicted staggered dimer AFM order as the magnetic order in ground state of the top-layer FeAs on As-terminated BaFe$_{2}$As$_{2}$(001), with a lower energy of 8 meV/Fe than the staggered trimer order, not only indicates that surface environment changes the magnetic ground state of the top-layer FeAs, but also shows that the As-terminated BaFe$_{2}$As$_{2}$(001) may serve as a good platform to discover novel magnetic order in experiment.

In the intermediate electron-doping region of the top-layer FeAs between the As termination and the full-Ba-coverage termination on BaFe$_{2}$As$_{2}$(001), the energy differences among different AFM orders of the top-layer FeAs are much reduced (Fig. 3). 
Actually, the diminished energy differences strongly enhance AFM spin fluctuations, while the enlarged energy differences with stable magnetic orders at two doping ends of the top-layer FeAs, i.e., the hole-doping end at As termination and the electron-doping end at Ba termination, suppress spin fluctuations. The variation tendency of energy differences (thus the strength of spin fluctuations) in the top-layer FeAs with doping concentrations is analogous to the dome of superconducting transition temperature $T_{c}$ in the experimental phase diagram of bulk Ba$_{1-x}$K$_x$Fe$_{2}$As$_{2}$ \cite{5HChen(2009)EPL}.
Our calculations indicate that surface doping can effectively tune the magnetic properties of the top-layer FeAs on BaFe$_2$As$_2$(001) and may probably induce a superconducting FeAs surface layer on its parent compound BaFe$_2$As$_2$. Experimentally, there are several evidences for surface doping effect on the superconductivity of FeSe on SrTiO$_3$ \cite{XueQK2012,ZhouXJ2012,FengDL2013,ZhouXJ2013,DingH2017,GuoJD2017} and FeSe on graphitized SiC(0001) \cite{SongCL2016}.

In addition to the impact of surface environment on the magnetic order of the top-layer FeAs, the magnetic order can in turn influence the STM image of BaFe$_{2}$As$_{2}$(001). In principle, the STM images are determined by local electronic structure rather than local atomic structure \cite{Binnig82APL,Binnig82PRL,Tersoff83}. Our above simulations show that in the staggered dimer AFM order of the top-layer FeAs on As termination, although surface As atoms are identical according to the atomic structure, they can be classified into two sublattices due to different electronic environments created by underlying Fe spins, finally giving out a $\sqrt{2}\times\sqrt{2}$ pattern. This pattern in alternating light and dark dots is different from the one of 50\%Ba-covered surface with surface Ba atoms arranged to the $\sqrt{2}\times\sqrt{2}$ square structure (Fig. 4). Previously, the $\sqrt{2}\times\sqrt{2}$ patterns have been reported by several STM experiments \cite{13MasseeF(2009)PRB,15NascimentoVB(2009)PRL,25HuiZhang(2010)PRB,Li2012PRB}. Thus the As termination with the top-layer FeAs in the staggered dimer AFM order should also be included when considering a $\sqrt{2}\times\sqrt{2}$ pattern.

In summary, we have investigated the magnetic properties and the STM images of BaFe$_{2}$As$_{2}$(001) surface at different terminations by using first-principles electronic structure calculations. For As termination, we find that the staggered dimer AFM order, instead of the collinear (single stripe) AFM order, is of the magnetic ground state of the top-layer FeAs. With increasing electron doping by depositing K/Ba adatoms, the magnetic ground state of the top-layer FeAs transforms to the collinear AFM order, mediating a strong-spin-fluctuation region with diminished energy differences among different AFM orders.
To identify the novel staggered dimer AFM order in the top-layer FeAs at the As termination, we have simulated its STM image, which shows a different $\sqrt{2}\times\sqrt{2}$ pattern from the 50\%Ba-covered case. These results reveal that the magnetic properties of the top-layer FeAs on BaFe$_{2}$As$_{2}$(001) can be tuned effectively by surface doping and call for special attention on the magnetic state at surface when explaining related STM images.

\begin{figure*}[!ht]
\includegraphics[angle=0,scale=0.5,clip=true]{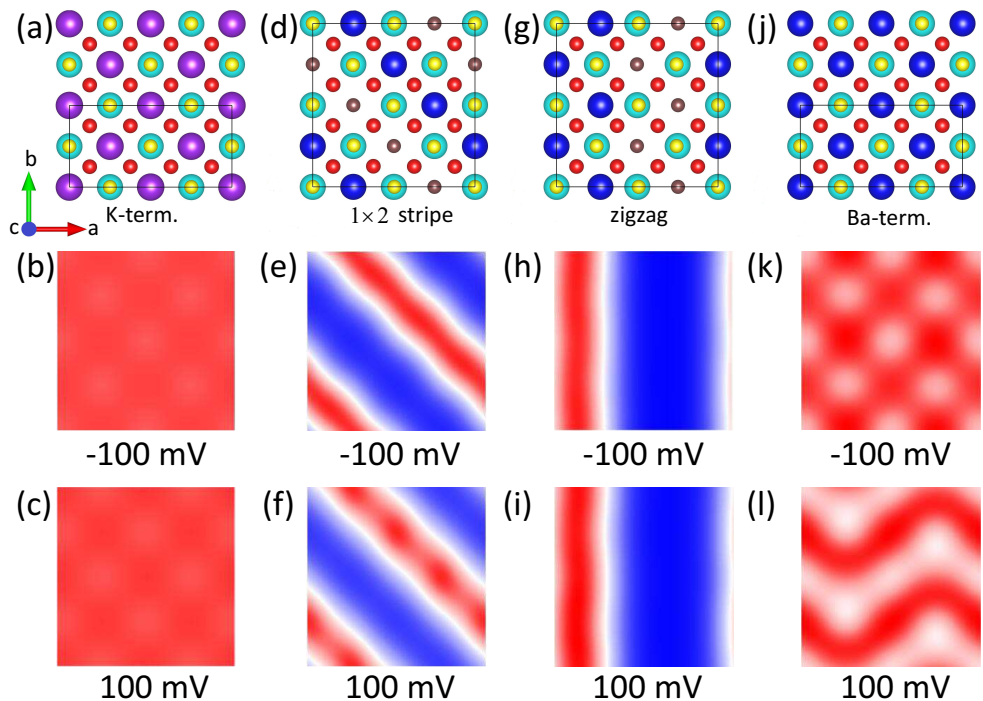}
\caption{(Color online) The simulated STM images of (a)-(c) K termination, 50\%Ba-covered terminations with surface Ba atoms in (d)-(f) the $1\times2$ stripe pattern and (g)-(i) the zigzag stripe pattern, and (j)-(l) Ba termination of BaFe$_{2}$As$_{2}$(001) with the top-layer FeAs in the collinear AFM order under biases of -100 mV and 100 mV.}
\label{fig5}
\end{figure*}

\begin{acknowledgments}

We wish to thank Miao Gao, Shu-Heng Pan, Xue-Jin Liang, and Jia-Xin Yin for helpful conversations. This work was supported by National Key R\&D Program of China (2017YFA0302903), National Natural Science Foundation of China (Grants No. 91421304 and No. 11474356), the Fundamental Research Funds for the Central Universities, and the Research Funds of Renmin University of China (RUC) (Grants No. 14XNLQ03 and No. 16XNLQ01). Computational resources were provided by the Physical Laboratory of High Performance Computing at RUC.

\end{acknowledgments}

\section{APPENDIX: SIMULATED STM IMAGES FOR OTHER TERMINATIONS of B\lowercase{a}F\lowercase{e}$_{2}$A\lowercase{s}$_{2}$(001)}

Figure 5 presents the simulated STM images for other terminations of BaFe$_{2}$As$_{2}$(001) surface with the top-layer FeAs adopting the collinear AFM order under both negative and positive biases. For the full-K-coverage termination [Fig. 5(a)], there is no apparent STM morphology [Figs. 5(b) and 5(c)]. For the 50\%Ba-covered case with surface Ba atoms arranged to the $1\times2$ stripe [Fig. 5(d)] and the zigzag stripe [Fig. 5(g)] structures respectively, their corresponding stripe patterns in STM images can be observed [Figs. 5(e), 5(f), 5(h), and 5(i)]. Instead, the full-Ba-coverage termination [Fig. 5(j)] demonstrates distinct images under different biases [Figs. 5(k) and 5(l)].

\end{document}